\documentclass[aps,twocolumn,superscriptaddress,altaffilletter,lengthcheck,
tightenlines,showpacs,showkeys]{revtex4}
\usepackage{amsmath} 
\usepackage{amssymb} 
\newcommand{\al}{\alpha}
\newcommand{\bb}{\beta}
\newcommand{\D}{\Delta}
\newcommand{\ben}{\begin{eqnarray}}
\newcommand{\een}{\end{eqnarray}}
\newcommand{\be}{\begin{equation}}
\newcommand{\ee}{\end{equation}}
\newcommand{\ba}{\begin{eqnarray}}
\newcommand{\ea}{\end{eqnarray}}
\newcommand{\n}{\label}

\newcommand{\ga}{\gamma}
\newcommand{\ro}{\rho}

\newcommand{\bn}{\begin{equation}\label}

\usepackage[dvipdf]{epsfig}
\usepackage{color}

\begin{document}

\title{Dark radiation and dark matter coupled to holographic Ricci dark energy}

\author{Luis P. Chimento}\email{chimento@df.uba.ar}
\affiliation{Departamento de F\'{\i}sica, Facultad de Ciencias Exactas y Naturales,  Universidad de Buenos Aires and IFIBA, CONICET, Ciudad Universitaria, Pabell\'on I, 1428, Buenos Aires  , Argentina}
\author{Mart\'{\i}n G. Richarte}\email{martin@df.uba.ar}
\affiliation{Departamento de F\'{\i}sica, Facultad de Ciencias Exactas y Naturales,  Universidad de Buenos Aires and IFIBA, CONICET, Ciudad Universitaria, Pabell\'on I, 1428, Buenos Aires  , Argentina}


\date{\today}
\bibliographystyle{plain}

\begin{abstract}

We investigate  a universe filled with interacting dark matter, holographic dark energy, and dark radiation for the spatially flat Friedmann-Robertson-Walker (FRW) spacetime. We  use a linear  interaction to  reconstruct all the component energy densities in terms of the scale factor by directly solving the balance's equations along with the source equation. We apply the $\chi^{2}$ method to the observational Hubble data for constraining the cosmic parameters, contrast with the Union 2 sample of supernovae, and analyze the amount of dark energy in the radiation era. It turns out that our model exhibits an excess of dark energy in the recombination era  whereas the stringent bound  $\Omega_{\rm x}(z\simeq 10^{10})<0.21$  at  big-bang nucleosynthesis is fulfilled. We find that the interaction provides a physical mechanism for  alleviating  the triple cosmic coincidence and this leads to  $\Omega_{\rm m0}/\Omega_{\rm x0} \simeq \Omega_{\rm r0}/\Omega_{\rm x0} \simeq {\cal O}(1)$. 

\end{abstract} 
\vskip 1cm

\keywords{ dark matter, holographic Ricci dark energy, dark radiation, linear  interaction}
\pacs{}

\bibliographystyle{plain}

\maketitle



\section{Introduction}
The two main components of the universe are dark matter and dark energy. The dark matter  accounts for $1/3$ of the stuff in the universe  and is  also inextricably connected with the formation of galaxies and galaxy clusters. In fact  this new form of matter not only holds galaxies together, but also is  responsible for the large-structure formation in the universe \cite{Book}.  The  astrophysical evidence  for dark matter come form colliding galaxies, gravitational lensing of mass distribution or power spectrum  of clustered matter \cite{dme}.  

The other $2/3$ exists as in an even more mysterious form dubbed as dark energy and is causing the expansion of the Universe to speed up, rather than slow down. The first serious  observational hints of dark energy in the universe date back to the 1990s when astronomers observations of supernova were used to trace the expansion history of the universe \cite{obse1}. Subsequently, the  WMAP results suggested that the aforesaid  amount of dark energy could  explain both the flatness of the universe along with the observed accelerated expansion  \cite{obse2}.  Nowadays, there is a growing number of observational methods for probing the dynamical behavior of dark energy at different scales; galaxy redshift surveys allow to obtain the Hubble expansion history by measurement of  baryon acoustic oscillation in the galaxy distribution \cite{Book}, geometric  weak lensing method applied to Hubble space telescope images helps to find tighter constraints on the  dark energy equation of state  also \cite{WL}.

Despite the devoted effort for understanding the nature of dark matter and dark energy,  there was not found  a microscopic theory for the dark side of the universe, capable of unraveling  their particles content. On the observational side, the data of several tests have confirmed some of their plausible properties, such as  that  it could be  the repulsive effect of dark energy or the clustering action of dark matter.  

Another interesting trait to explore is related with the existence of  non-gravitational coupling between dark matter and dark energy,  such exchange of energy  could alter the cosmic history, leaving testable imprints in the universe \cite{jefe1}.  It is believed  that a coupling between dark energy and dark matter changes the background evolution of the dark sector allowing to constrain any type of interaction and giving  rise to a richer cosmological dynamics compared with non-interacting models \cite{jefe1}. A step forward for constraining dark matter and dark energy  is to use  the physics behind  recombination or big-bang nucleosynthesis epochs by adding a decoupled radiation term to the dark sector for taking into account the stringent bounds  related to  the behavior of dark energy at early times \cite{hmi1}, \cite{hmi2}. The behavior of dark energy in the recombination era was explored within the framework of three interacting components also \cite{jefe2}.

Our goal is to consider  a model where dark matter and  dark radiation are coupled to holographic dark energy  and explore the cosmic triple coincidence problem related to the amount of these components at present \cite{TCC}. We  perform a cosmological constraint using the updated Hubble data \cite{H19}, numerically obtain the  distance modulus $\mu(z)$ for contrasting with  the Union 2 compilation   of supernovae Ia \cite{amanu}, and analyze the order of magnitude corresponding to  the cosmological parameter known as transition redshift. In order to check the feasibility of the model, we also examine the severe bounds for  dark energy in the recombination era \cite{EDE1} or nucleosynthesis epoch \cite{Cyburt}.   



\section{The model }
We consider a spatially flat homogeneous  and isotropic universe described by FRW spacetime with line element given by  $ds^{2}=-dt^{2}+a^{2}(t) (dx^{2}+dy^{2}+dz^{2})$ being  $a(t)$ the scale factor. The universe is  filled with three interacting fluids namely, dark radiation, dark matter and modified holographic Ricci dark energy  so that  the evolution of the FRW universe is governed by  the Friedmann and conservation equations, respectively,  
\be
\n{01}
3H^{2}=\ro_{\rm r}+\ro_{\rm m}+\ro_{\rm x},
\ee
\be
\n{02}
\dot{\ro}+3H(\ro_{\rm r}+p_{\rm r}+\ro_{\rm m}+p_{\rm m}+\ro_{\rm x}+p_{\rm x})=0,
\ee
where $a$  is the scale factor and $H = \dot a/a$ stands for the Hubble expansion rate. Here, we will use the holographic principle within the cosmological context  by associating the infrared cutoff $L$ with the dark energy density, thus we take $L^{-2}$ in the form of a linear combination of $\dot{H}$ and $H^{2}$:
\be
\n{03}
\ro_{\rm x}=\frac{2}{\al -\bb}\left(\dot H + \frac{3\al}{2} H^2\right),
\ee
Here,  $\al$ and $\bb$ are  two free constants. In particular, we obtain $\ro_{\rm x}\propto R$ for  $\al=4/3$, where $R=6(\dot{H}+2H^{2})$ is the Ricci scalar curvature for a spatially flat FRW space-time.

The use of the variable $\eta = \ln(a/a_0)^{3}$, where $a_0$ is set as the value of the scale factor at present and $'\equiv d/d\eta$, allows us to rewrite Eqs. (\ref{02}) and (\ref{03}) as
\be
\n{4}
\ro'=-\ga_{\rm r}\ro_{\rm r}-\gamma_{\rm m}\ro_{\rm m}-\gamma_{\rm x}\ro_{\rm x} ,
\ee
\bn{5}
\ro'=-\al\ro_{\rm c}-\bb\ro_{\rm x},
\ee
\bn{55}
\ro_{\rm c}=\ro_{\rm r}+\ro_{\rm m},
\ee
where $\gamma_{i}=1+p_{i}/\ro_{i}$ denotes  the barotropic index, not necessarily constant, of each component with $i=\{r,m,x\}$, $\ga_{\rm r}\simeq 4/3$, $\ga_{\rm m}\simeq 1$, $0<\ga_{\rm x}<2/3$ so that $0<\ga_{\rm x}<\ga_{\rm m}<\ga_{\rm r}$.  Taking into account  Eq. (\ref{5}) along with  Eq. (\ref{55}), we could extract as a physical hint that the modified holographic dark energy  (\ref{03}) forces to the dark matter and dark radiation to have the same bare equation of state.

The holographic dark energy (\ref{03}) or (\ref{5}), looks like  a ``conservation equation" for the three dark components with constant coefficients. Therefore, the selected holographic dark energy (\ref{03}) or (\ref{5}) has transformed the original model of three interacting components into another simpler scheme  of  two  interacting components having two constant equations of state . As we have already mentioned above, such degeneration occurs because the dark radiation and dark matter have the same equation of state, $p_{\rm r}=(\al-1)\ro_{\rm r}$ and $p_{\rm m}=(\al-1)\ro_{\rm m}$ . After comparing the whole conservation equation  (\ref{4}) with modified conservation equation  (\ref{5}),  we obtain the compatibility relation  
\be
\n{07}
\ga_{\rm r}\ro_{\rm r}+\gamma_{\rm m}\ro_{\rm m}+\gamma_{\rm x}\ro_{\rm x}=\al(\ro_{\rm r}+\ro_{\rm m})+\bb\ro_{\rm x},
\ee
that relates the equation of state of the dark components with the bare ones. In what follows, we will use Eq. (\ref{5}) with constant coefficients $\al$ and $\beta$ instead of  Eq.  (\ref{4}) with  the non-constant coefficient $\ga_{\rm x}=1+p_{\rm x}/\ro_{\rm x}$.  In some sense,  (\ref{4}) and  (\ref{5}) give rise to  different representations of the mixture of two interacting fluids and clearly these descriptions are related between them by the compatibility relation (\ref{07}). Therefore, the holographic dark energy  (\ref{03}) conveniently links a model of three interacting fluids having {\bf variable equations of state} with a model of two interacting fluids with ``{\bf bare constant equations of state}". 

Solving the system of  equations $\ro=\ro_{\rm c}+\ro_{\rm x}$ and (\ref{5}) we get the energy densities of both component as a function of $\ro$ and its derivative $\ro'$ 
\bn{cx}
\ro_{\rm c}=-\frac{\bb\ro+\ro'}{\D\,}, \qquad \ro_{\rm x}=\frac{\al\ro+\ro'}{\D},
\ee
where $\D=\al-\bb$ is the determinant of the linear equation system and we  assume that $\bb<\al$. Now, we introduce  energy transfer between those components by  splitting  Eq. (\ref{5}) into two balance equations
\bn{cq}
\ro_{\rm c}'+\al\ro_{\rm c} =-Q.
\ee
\bn{xq}
\ro_{\rm x}'+\bb\ro_{\rm x} =Q,
\ee
where we have considered a coupling with a factorized $H$ dependence $3HQ$, being $Q$ the interaction term that generates the energy transfer between the two fluids. After differentiating the first Eq. (\ref{cx}) and combining with  Eq. (\ref{cq}), we obtain a second order differential equation for the total energy density (source equation) 
\be
\n{2} 
\rho''+(\al+\bb)\rho'+\al\bb\rho= Q\D.
\ee

We will take into account interactions $Q$ for which the solutions of the evolution equation for the scale factor $H=\sqrt{\ro/3}$ includes power law ones, because they play an essential role for determining the asymptotic behavior of the effective  barotropic index $\ga=(\al\ro_{\rm c}+\bb\ro_{\rm x})/\ro=-2\dot{H}/3H^{2}$. It describes a universe approaching to a stationary stage $\ga_s$ associated with the constant solution $\ga=\ga_s$, where $0<\bb<\ga_s<\al$ and $a=t^{2/3\ga_{s}}$. In addition, given a set of initial conditions, if  $\ga$ tends asymptotically to the  constant solution $\ga_s$, then $\ga_{s}$ becomes an attractor solution \cite{jefe1}. An interaction satisfying this requirement belongs to the class 
\bn{qe}
Q=\frac{(\ga_s-\al)(\ga_s-\bb)}{\D}\,\ro,
\ee
with $Q<0$ \cite{jefe1}. Solving the  source equation (\ref{2}) for this interaction, we obtain the total energy density  
\be
\n{441}
\ro=c\,a^{-3\ga_s}+b\,a^{-3(\al+\bb-\ga_s)}.
\ee
Hence, for any initial conditions $c$, $b$,  and  large scale factor,  the energy density behaves as $ c/a^{3\ga_s}$ and the power-law expansion $a\to t^{2/3\ga_s}$ becomes asymptotically stable. 

The  dark densities of coupled components, $\ro_{\rm x}$ and $\ro_{\rm c}$,  are given by 
\be
\n{re1}
\ro_{\rm c}=\frac{\left(\ga_s-\bb\right)c\,a^{-3\ga_s}+\left(\al-\ga_s\right)b\,a^{-3(\al+\bb-\ga_s)}}{\D},
\ee
\be
\n{re21}
\ro_{\rm x}=\frac{\left(\al-\ga_s\right)c\,a^{-3\ga_s}+\left(\ga_s-\bb\right)b\,a^{-3(\al+\bb-\ga_s)}}{\D}.
\ee
Thus, the ratio $r=\ro_{\rm c}/\ro_{\rm x}$ tends to $r_{\rm s}= (\ga_s-\bb)/(\al-\ga_s)$, being $r_{\rm s}$ an attractor.

With the aid of the total energy density (\ref{441}), we can calculate the explicit form of the interaction term (\ref{qe}) as a function of the scale factor, 
\bn{q}
Q=\frac{(\ga_s-\al)(\ga_s-\bb)}{\D}\left[c\,a^{-3\ga_s}+b\,a^{-3(\al+\bb-\ga_s)}\right].
\ee
In turn  Eq. (\ref{cq}) can be rewritten as
\bn{cqq}
\ro_{\rm r}'+\ro_{\rm m}'+\al(\ro_{\rm r}+\ro_{\rm m})=-Q.
\ee
In order to  break the degeneracy of this set of components, formed by dark radiation and dark matter, we introduce partial interactions into the corresponding balance equation  of both components as follows: 
\bn{06}
\ro_{\rm r}'+\al\ro_{\rm r}=Q_{\rm r}.
\ee
\bn{05}
\ro_{\rm m}'+\al\ro_{\rm m}=Q_{\rm m},
\ee
where $Q_{\rm r}$ and $Q_{\rm m}$ stand for  the exchange of energy  between $\ro_{\rm r}$ and $\ro_{\rm x}$, and besides these satisfy the condition
\be
\n{06b}
 Q+Q_{\rm m}+Q_{\rm r}=0,
\ee
to recover the conservation equation  (\ref{5}) after having summed all Eqs.  (\ref{cqq})--(\ref{05}). Also we assume that $Q_r$ and $Q_m$ are a linear combination of the two terms contained in the interaction term $Q$ (\ref{q}), so they read
\bn{qr}
Q_{\rm r}=c_1\,a^{-3\ga_s}+b_1\,a^{-3(\al+\bb-\ga_s)},
\ee
\bn{qm}
Q_{\rm m}=-(c_1+ Q_{0}c)\,a^{-3\ga_s}-(b_1+ Q_{0} b)\,a^{-3(\al+\bb-\ga_s)},
\ee
where $Q_{0}=(\ga_s-\al)(\ga_s-\bb)/\D$,   while $c_1$ and $b_1$ are free parameters of the model. Inserting these interactions into the evolution equation of the dark radiation and dark matter (\ref{06})--(\ref{05}) and solving these coupled system of equations we obtain
\bn{rr}
\ro_{\rm r}=\frac{c_1\,a^{-3\ga_s}}{\al-\ga_s}-c_0\,a^{-3\al}+\frac{b_1\,a^{-3(\al+\bb-\ga_s)}}{\ga_s-\bb},
\ee
\bn{rm}
\ro_{\rm m}=-\frac{(c_1+ Q_{0}c)\,a^{-3\ga_s}}{\al-\ga_s}+c_0\,a^{-3\al}-\frac{(b_1+ Q_{0}b)\,a^{-3(\al+\bb-\ga_s)}}{\ga_s-\bb},
\ee
where the first and third terms in both dark energies densities are the particular solutions of the evolution equations (\ref{06}) and (\ref{05}) while the second term, in both Eqs. (\ref{rr}) and (\ref{rm}), is the homogeneous solution of the linear system of equations (\ref{06}) and (\ref{05}). In what follows, we fix $c_{1}=-Q_{0}c/2>0$ and $b_{1}=-Q_{0}b/2>0$ without loss of generality. Using Eq. (\ref{re1}), (\ref{re21}), (\ref{rr}), and (\ref{rm}), we find the coefficients $c$, $b$, and $c_{0}$ in terms of the the density parameters and $\ga_{s}$. Now, we only show their expressions for  $\al=4/3$ and $\al +\beta-\ga_{s}=1$:

\begin{eqnarray}
\label {c}
c&=&3H^{2}_{0}\left[\frac{1+ \Omega_{\rm x0}(3\ga_{s}-5)}{3(\ga_{s}-1)}\right],\
\\
\label {b}
b&=&3H^{2}_{0}\left[\frac{(5 \Omega_{\rm x0}-4) +3\ga_{s}(1- \Omega_{\rm x0})}{3(\ga_{s}-1)}\right],\
\\
\label{c0}
c_{0}&=&3H^{2}_{0}\left[\frac{1- \Omega_{\rm x0}-2\Omega_{\rm m0}}{2}\right].
\end{eqnarray}
Here, $\Omega_{\rm i0}$ stands for the density parameter of  each component.

\section{Observational constraints on the three interacting  model}
We will provide a qualitative estimation of the cosmological parameters by constraining them with the Hubble data  \cite{obs3}- \cite{obs4} and the strict bounds for the behavior of dark energy at early times \cite{EDE1}. In the former case, the  statistical analysis is based on the $\chi^{2}$--function of the Hubble data which is constructed as (e.g.\cite{Press})
\be
\n{c1}
\chi^2(\theta) =\sum_{k=1}^{19}\frac{[H(\theta,z_k) - H_{\rm obs}(z_k)]^2}{\sigma(z_k)^2},
\ee
where $\theta$ stands for cosmological parameters, $H_{\rm obs}(z_k)$ is the observational $H(z)$ data at the redshift $z_k$, $\sigma(z_k)$ is the corresponding $1\sigma$ uncertainty, and the summation is over the $19~$ observational  $H(z)$ data \cite{H19}.  Using the absolute ages of passively evolving galaxies observed at different redshifts, one obtains the differential ages $dz/dt$ and the function $H(z)$ can be measured through the relation $H(z)=-(1+z)^{-1}dz/dt$ \cite{obs3}, \cite{obs4}. The data  $H_{\rm obs}(z_i)$ and $H_{\rm obs}(z_k)$ are uncorrelated because they were obtained from the observations of galaxies at different redshifts. 

From Eq. (\ref{441}) one finds that the Hubble expansion of the model  becomes
\be
\n{Ht}
H(\theta| z)=H_{0} \Big( {\cal C}x^{3\ga_{s}}+{\cal B}x^{3(\al +\beta-\ga_{s})}\Big)^{\frac{1}{2}}
\ee
where  $c=3H^{2}_{0}{\cal C}$,  $b=3H^{2}_{0}{\cal B}$ as  obtained form (\ref{c}), (\ref{b}),respectively. Here, we consider  $\theta=\{H_{0},\ga_{\rm s},  \Omega_{\rm  x0}\}$ as the independent parameters to be constrained  for the model encoded in the Hubble function (\ref{Ht}) with the statistical estimator (\ref{c1}), while $\alpha$ is taken equal to $4/3$ to have an early era dominated by radiation. Besides, we will impose  $\al +\beta-\ga_{s}=1$ so that the universe exhibits an intermediate stage dominated by pressureless dark  matter.   For a given pair of $(\theta_{1}, \theta_{2})$,  we are going to perform the statistic analysis by minimizing the $\chi^2$ function to  obtain the best fit values of  the random variables $\theta_{\rm crit}=\{\theta_{\rm crit 1}, \theta_{\rm crit 2}\}$ which correspond to a minimum of Eq.(\ref{c1}). Then, the  best-fit parameters $\theta_{\rm crit}$ are those values where $\chi^2_{\rm min}(\theta_{\rm crit})$ leads to the local minimum of the $\chi^2(\theta)$ distribution. If $\chi^2_{\rm d.o.f}=\chi^2_{\rm min}(\theta_{\rm crit})/(N -n) \leq 1$ the fit is good and the data are consistent with the considered model $H(z;\theta)$. Here, $N$ is the number of data and $n$ is the number of parameters \cite{Press}. The variable $\chi^2$ is a random variable that depends on $N$ and its probability distribution is a $\chi^2$ distribution for $N-n$ degrees of freedom.  Here $N=19$ and $n=3$, so in principle, we will perform  three minimizations of the  $\chi^{2}$ statistical estimator, interpreting the goodness of  fit by checking the condition $\chi^2_{\rm d.o.f}<1$; as a way to keep the things clear and focus on extracting relevant physical information from this statistical estimation.  

\begin{figure}[hbt!]
\begin{minipage}{1\linewidth}
\resizebox{1.6in}{!}{\includegraphics{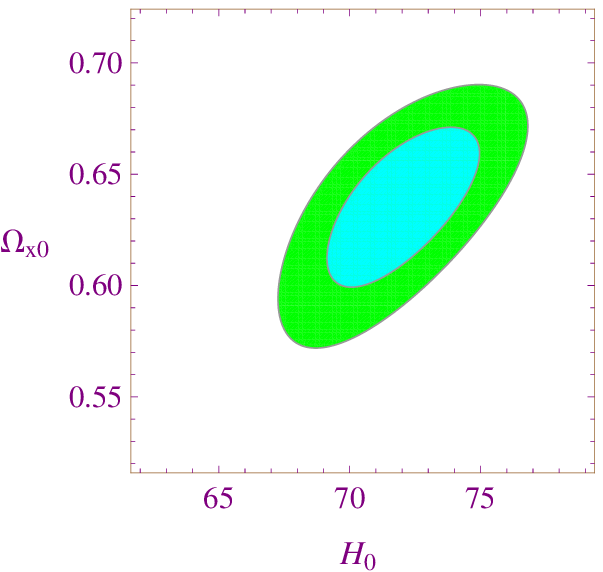}}\hskip0.05cm
\resizebox{1.6in}{!}{\includegraphics{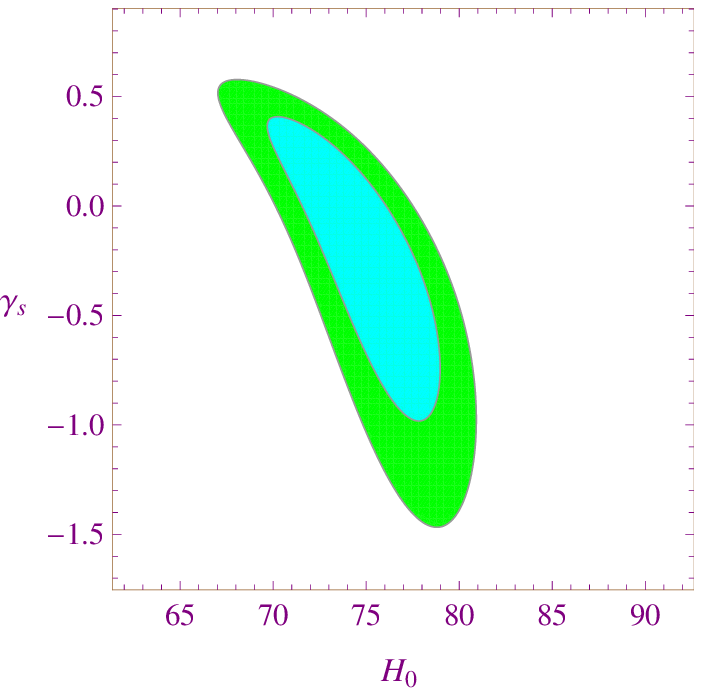}}\hskip0.05cm
\resizebox{1.6in}{!}{\includegraphics{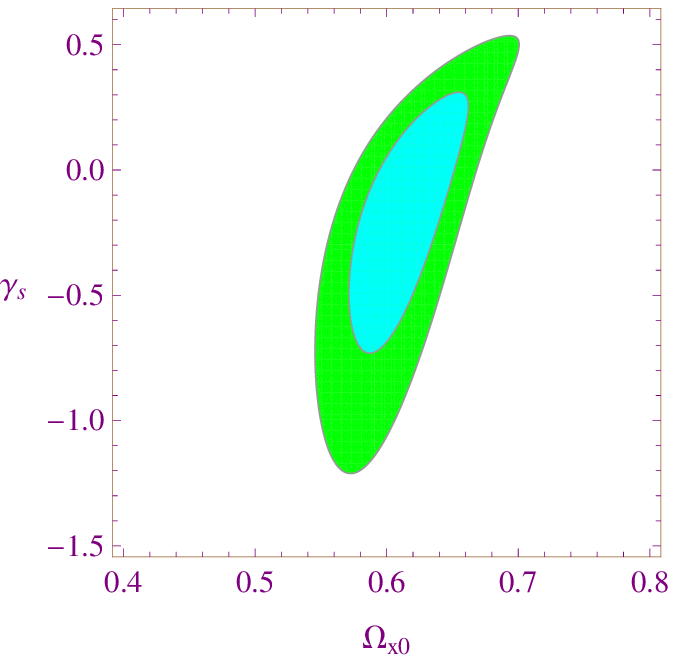}}\hskip0.05cm 
\caption{\scriptsize{Two-dimensional C.L. associated with $1\sigma$,$2\sigma$ for different $\theta$ planes.}}
\label{Fig1}
\end{minipage}
\end{figure}

\begin{figure}[hbt!]
\begin{minipage}{1\linewidth}
\resizebox{3.5in}{!}{\includegraphics{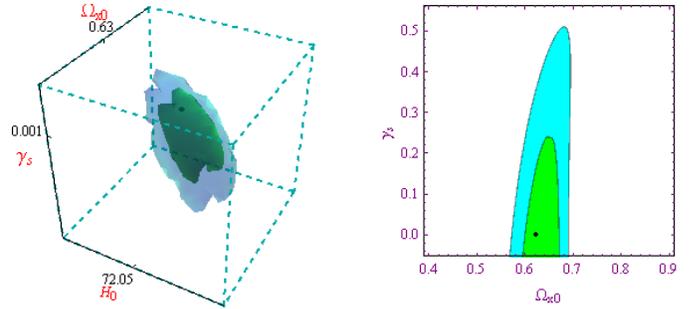}}\hskip0.09cm
\caption{\scriptsize{ (Left panel):Three-dimensional C.L. for  the $H_{0}-\gamma_{s}-\Omega_{\rm x0}$ plane.  (Right panel): Two-dimensional C.L. obtained after have performed the marginalization over $H_{0}$. }}
\label{Fig2}
\end{minipage}
\end{figure}

\begin{figure}[hbt!]
\begin{minipage}{1\linewidth}
\resizebox{1.6in}{!}{\includegraphics{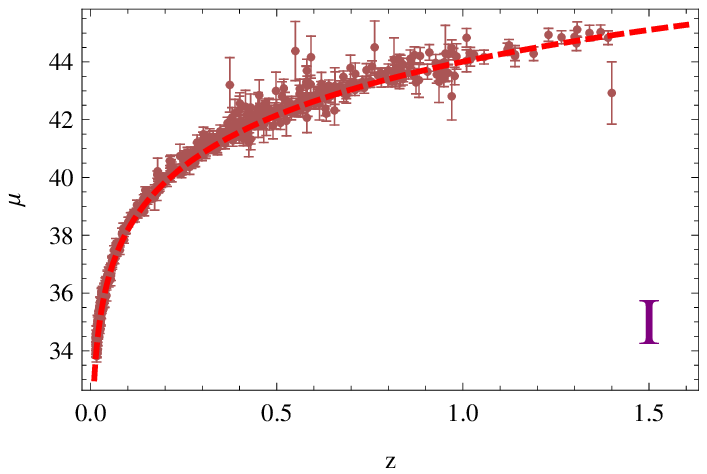}}\hskip0.03cm
\resizebox{1.6in}{!}{\includegraphics{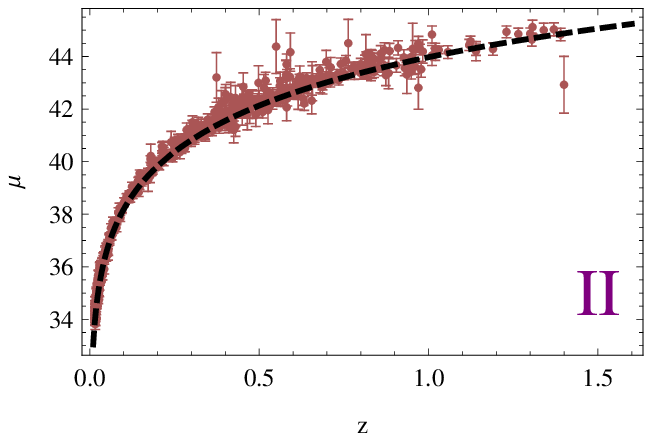}}\hskip0.03cm
\resizebox{1.6in}{!}{\includegraphics{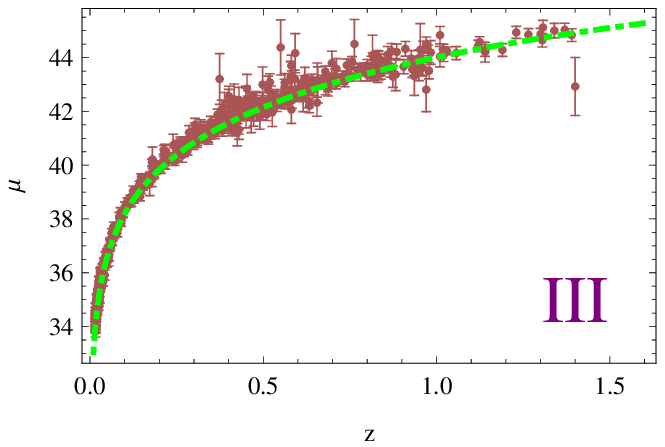}}\hskip0.03cm
\resizebox{1.6in}{!}{\includegraphics{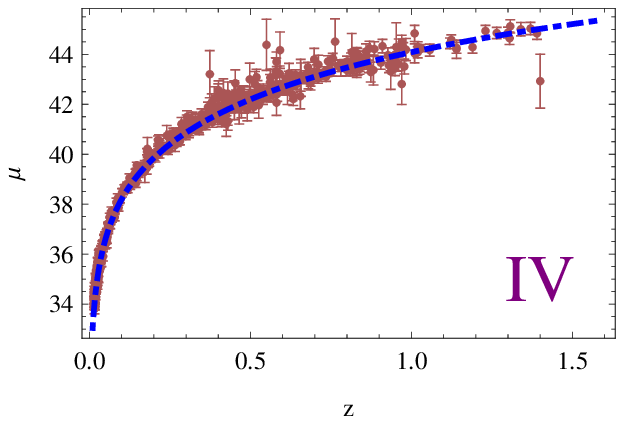}}\hskip0.03cm
\resizebox{1.65in}{!}{\includegraphics{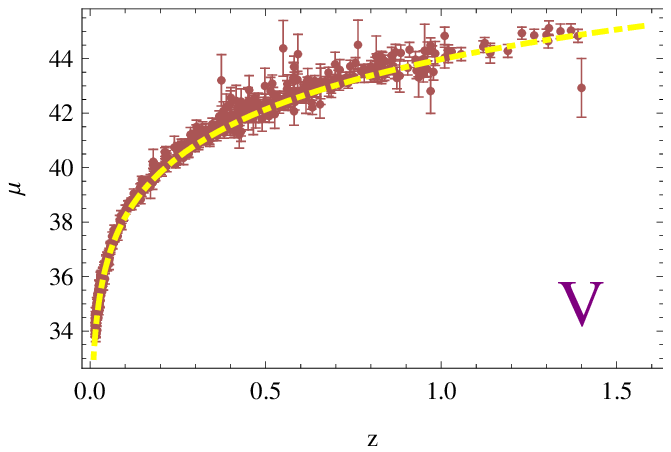}}\hskip0.07cm
\caption{\scriptsize{Hubble diagram for the Union2 compilation: the points are the observational data of  supernovae, the theoretical curves $\mu(z)$ (dashed lines) represent the best cosmological model for different cases: (global estimation) $~I-(H_{0}, \Omega_{\rm x0}, \ga_{s})=(72.05, 0.63, 0.001)$, (marginalizing over $H_{0}$) $~II-(\Omega_{\rm x0}, \ga_{s})=(0.619^{+0.052}_{-0.023}, 0.001^{+0.239}_{-0.239})$, $III- (H_{0}, \Omega_{\rm x 0})= (72.05^{+2.90}_{-2.93}, 0.637^{+0.033}_{-0.038})$, $IV-( H_{0},\ga_{s})=(73.81^{+5.14}_{-4.24}, 0.001^{+0.404}_{-0.984})$, $V-(\Omega_{\rm x 0},\ga_{s})=(0.623^{+0.038}_{-0.051},0.001^{+0.310}_{-0.732})$.}}
\label{Fig5}
\end{minipage}
\end{figure}

\begin{figure}[hbt!]
\begin{minipage}{1\linewidth}
\resizebox{3.5in}{!}{\includegraphics{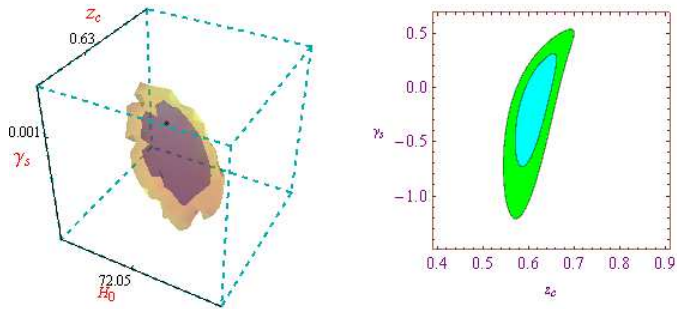}}
\caption{\scriptsize(Upper panel):Three-dimensional C.L. for  the $H_{0}--\gamma_{s}--z_{c}$ plane.  (Lower panel): Two-dimensional C.L. obtained after have performed the marginalization over $H_{0}$.}
\label{Fig3}
\end{minipage}
\end{figure}

\begin{figure}[hbt!]
\begin{minipage}{1\linewidth}
\resizebox{1.6in}{!}{\includegraphics{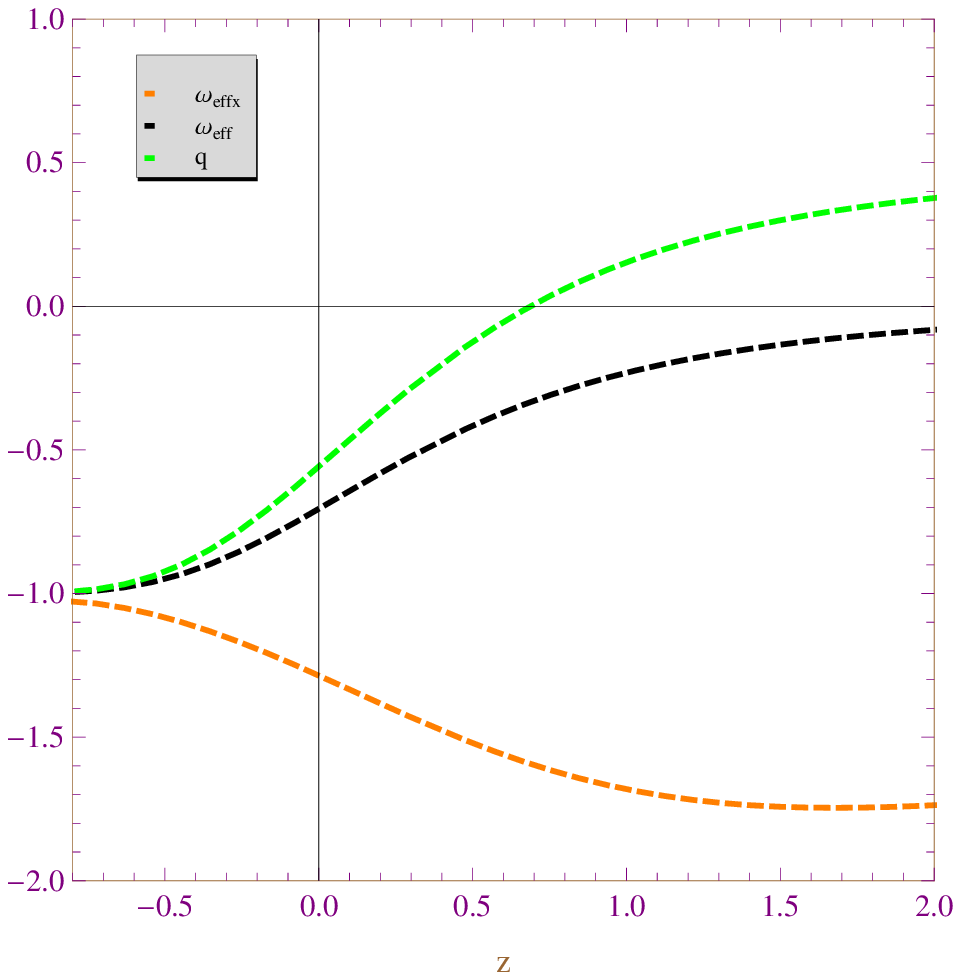}}
\resizebox{1.6in}{!}{\includegraphics{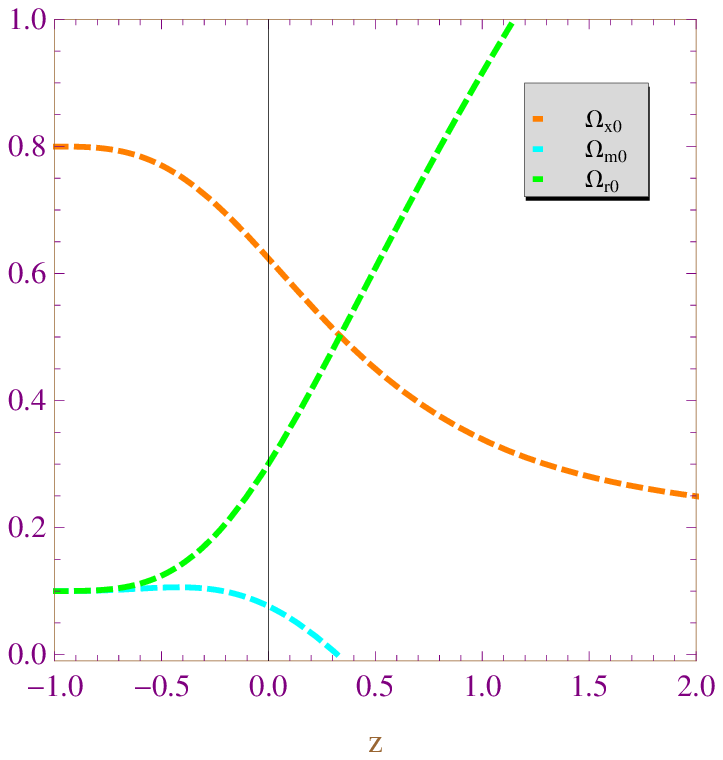}}
\resizebox{1.73in}{!}{\includegraphics{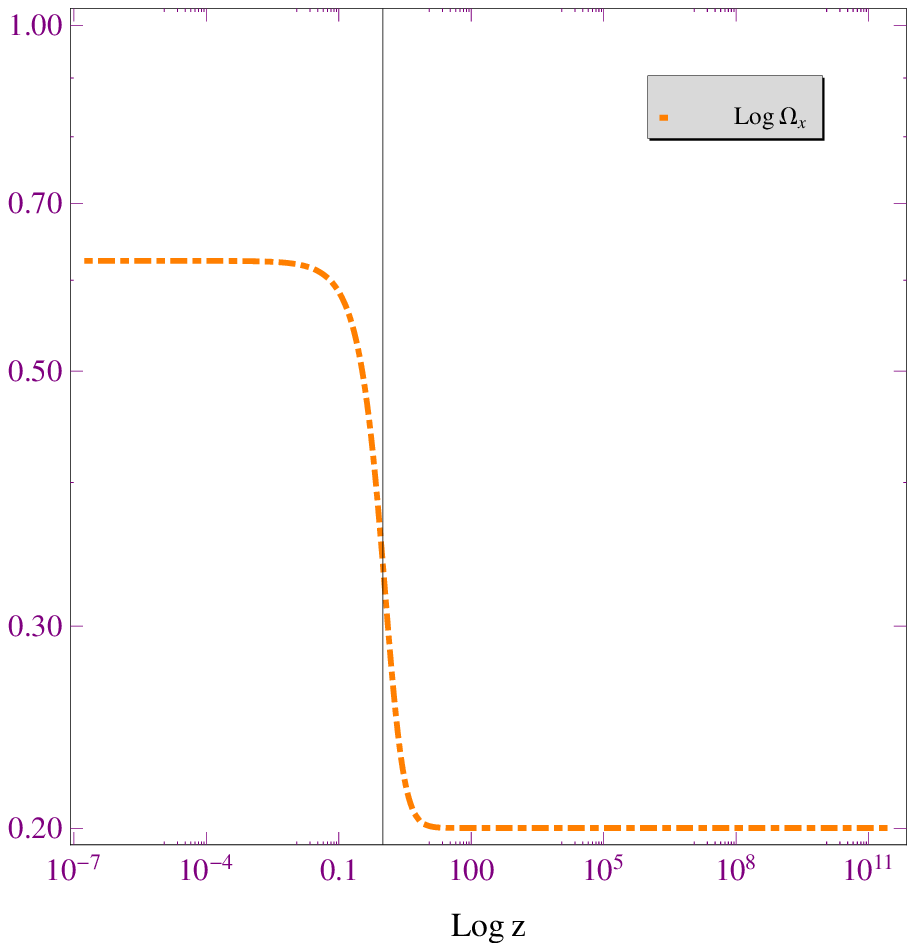}}
\caption{\scriptsize{Plots of $\Omega_{\rm x}(z)$,  $\Omega_{\rm m}(z)$,  $\Omega_{\rm r}(z)$,  $q(z)$, $\omega_{\rm eff}(z)$, $\omega_{\rm effx}(z)$ using the best-fit values  obtained  with the Hubble data for different $\theta$ planes. Plot of  ${\rm Log}\Omega_{\rm x}$ in terms of ${\rm Log }~z$.}}
\label{Fig4}
\end{minipage}
\end{figure}

\begin{center}
\begin{table}
\begin{minipage}{1\linewidth}
\scalebox{0.9}{
\begin{tabular}{|l|l|l|l|l|l|l|l|l|l|l|}
\hline
\multicolumn{3}{|c|}{2D Confidence level} \\
\hline
Priors & Best fits& $\chi^{2}_{\rm d.o.f}$ \\
\hline
{ $\ga_{s}=10^{-3}$}& ($H_{0}, \Omega_{\rm x 0})= (72.05^{+2.90}_{-2.93}, 0.637^{+0.033}_{-0.038})$& $0.81$\\
{$\Omega_{\rm x 0}=0.67$}&$( H_{0},\ga_{s})=(73.81^{+5.14}_{-4.24}, 0.001^{+0.404}_{-0.984})$& $0.94$\\
{$ H_{0}=70.4$}&$(\Omega_{\rm x 0},\ga_{s})=(0.623^{+0.038}_{-0.051},0.001^{+0.310}_{-0.732})$& $0.86$\\
\hline
\end{tabular}}
\caption{\scriptsize{ Observational bounds for the 2D C.L. obtained in  Fig. (\ref{Fig1}) by varying two cosmological parameters. The $\chi^{2}_{\rm d.o.f}$ in all the cases studied is less the unity; in fact it goes from 0.81 to 0.94.}}
\label{I}
\end{minipage}
\end{table}
\end{center}

The  random datasets that satisfy the inequality $\Delta\chi^{2}=\chi^2(\theta)-\chi^{2}_{\rm min}(\theta_{\rm crit})\leq 2.30$ also represent $68.3\%$ confidence  contours in the 2D plane at $1\sigma$ level. It can be shown that $95.4\%$ confidence contours  with a $2\sigma$  error bar in the samples satisfy $\Delta\chi^{2}\leq 6.17$. The two-dimensional C.L. obtained with the standard $\chi^{2}$ function
for two independent parameters is shown in Fig. (\ref{Fig1}), whereas the estimation
of these cosmic parameters is briefly summarized in Table (\ref{I}).

 We obtain
$\ga_{s} \leq 10^{-3} $, so  these values clearly fulfill
the constraint $\ga_{s}<2/3$ which ensures the existence of an accelerated phase of
the universe at late times [Table (\ref{I})]. We find the best fit at 
$(H_{0}, \Omega_{\rm x 0})= (72.05^{+2.90}_{-2.93} {\rm km~s^{-1}\,Mpc^{-1}}, 0.637^{+0.033}_{-0.038})$ with $\chi^{2}_{\rm d.o.f}=0.81$
by using the prior $ \ga_{s}=10^{-3}$. These findings show, in broad terms,  
that the  estimated values of $H_{0}$ and $\Omega_{\rm x 0}$ are in agreement with 
the standard ones reported by the WMAP-7 project \cite{WMAP7}. The value of $\Omega_{\rm x 0}$
is slightly lower than the standard one of $0.7$ being such discrepancy less or equal to  $0.1\%$.
We find that using the priors  $H_{0}=70.4{\rm km~s^{-1}\,Mpc^{-1}}$
the best-fit values for the present-day density parameters are   
$(\Omega_{\rm x 0},\Omega_{\rm r0})=(0.62,   7\times 10^{-2})$ along with a larger goodness condition ($\chi^{2}_{\rm d.o.f}=0.86$) [Table (\ref{I})]. 
Regarding  the amount of  dark matter at present,  we  have  fixed  $\Omega_{m0}=0.3$  because this value is consistent with one reported by the WMAP-7 project \cite{WMAP7}. In 
performing the statistical analysis, we find that $H_{0} \in [70.4, 73.81]{\rm km~s^{-1}\,Mpc^{-1}}$ 
so the estimated values are met  within $1 \sigma$ C.L.  reported
by Riess \emph{et al} \cite{H0}, to wit,  $H_{0}=(72.2 \pm 3.6){\rm km~s^{-1}\,Mpc^{-1}}$. 
In order to ensure that our previous local minimization analysis is  correct, we have performed a global statistical analysis by estimating all the parameters at once. In doing that, we obtain $(H_{0}, \Omega_{\rm x0}, \ga_{s})=(72.05, 0.63, 10^{-3})$ along with $\chi^{2}_{\rm d.o.f}=0.86<1$ [see Fig. (\ref{Fig2})]. Fig.  (\ref{Fig2}) shows  two-dimensional C.L in the 
$\Omega_{\rm x0}-- \ga_{s}$ plane obtained when  the joint probability $P(H_{0}, \Omega_{\rm x0}, \ga_{s})$  is  marginalized over $H_{0}$ (see Fig.  (\ref{Fig2})); then the marginalized  best-fit values  become  $(\Omega_{\rm x0}, \ga_{s})=(0.619^{+0.052}_{-0.023}, 0.001^{+0.239}_{-0.239})$  with $\chi^{2}_{\rm d.o.f}=0.87<1$. It must be stressed that  we report for the most relevant  minimization procedures the corresponding marginal $1\sigma$ error bars \cite{Bayes} as can be  seen in Table (\ref{I}).

As is well known, distance indicators can be used for confronting distance measurements to the corresponding
model predictions. Among the most useful ones are those objects of known
intrinsic luminosity such as  standard candles, so that the corresponding comoving
distance can be determined. That way, it is possible to reconstruct the Hubble expansion rate by searching this
sort of object at different redshifts. The most important
class of such indicators is type Ia supernovae.  Then, we would like to   compare the Hubble data  with the
Union 2 compilation of 557 SNe Ia \cite{amanu} by contrasting theoretical distance modulus with the observational dataset. In order to do that, we note that the apparent magnitude of a supernova placed at a given
redshift $z$ is related to the expansion history of the Universe through the distance modulus
\be
\n{mu}
\mu\equiv m -M= 5\log \frac{d_{L}(z)}{h}+\mu_{0},
\ee
where $m$ and $M$ are the apparent and absolute magnitudes, respectively, $\mu_{0}=42.38$, $h=H_{0}/100\rm{km^{-1} s^{-1}}$, and $d_{L}(z)=H_{0}(1+z)r(z)$, $r(z)$  being the comoving distance, given for a FRW metric by 
\be
\n{r}
r(z)=\int^{z}_{0}{\frac{dz'}{H(z')}}
\ee
Using the Union 2 dataset, we will obtain five Hubble diagrams and compare each of them with the theoretical  distance modulus curves  that represent the best-fit cosmological models found with the update Hubble data (see Fig. (\ref{Fig5})); it turned out that at low redshift ($z<1.4$) there is an excellent  agreement between the theoretical model and the observational data.

For the sake of completeness,
we also report bounds for other cosmological relevant parameters [see Table (\ref{II})], such as the fraction of dark radiation $\Omega_{\rm r}(z=0)$, the effective  equation of state at $z=0$ ($\omega_{\rm eff0}=\ga_{\rm eff0}-1$), decelerating parameter  at the present time $q_{0}$, and the transition redshift ($z_{\rm c}$) among many others, all these quantities are derived
using the three best fit values reported in Table (\ref{I}). We find that the $z_{c}$ is of the order unity varying over the interval $[0.68, 0.93]$, such values are close to $z_{\rm c}=0.69^{+0.20}_{-0.13}$ reported in \cite{Zt1}, \cite{Ztn}  quite recently. Moreover, taking into account
a $\chi^{2}$-statistical analysis made in the $(\omega_{\rm 0}, z_{c})$-plane based on the supernova sample 
(Union 2) it has been  shown that at  $2 \sigma$ C.L.  
the transition redshift  varies from $0.60$ to $1.18$ \cite{Zt2}. In order to estimate $z_{c}$ as independent parameter using a $\chi^{2}$ method, we  first needed to obtain  its generic formula by imposing the condition $(z_{\rm c}+1)H'(z_{\rm c})=H(z_{\rm c})$:
\be
\label{zc}
 z_{c}=\left(\frac{3\Omega_{\rm x0}(1-\ga_{s})+ (4-3\ga_{s})}{(2-3\ga_{s})[3\Omega_{\rm x0}(1-\ga_{s})+ 1]}\right)^{1/3(1-\ga_{s})}-1. 
\ee
Hence, placing Eq. (\ref{zc}) into Eq. (\ref{Ht}), the Hubble parameter turns out to be  a function of  $H_{0}$, $z_{\rm c}$, and $\ga_{s}$. The global statistical analysis using $z_{c}$ as independent parameter instead of $\Omega_{x0}$ leads to  $(H_{0}, z_{\rm c}, \ga_{s})=(72.05, 0.63, 10^{-3})$ along with $\chi^{2}_{\rm d.o.f}=0.86$ [see Fig. (\ref{Fig3})] whereas the marginalized best-fit values  become  $(z_{\rm c}, \ga_{s})=(0.623^{+0.039}_{-0.052}, 0.001^{+0.313}_{-0.733})$  with a $\chi^{2}_{\rm d.o.f}=0.86$ [see Fig. (\ref{Fig3})]; notice that the marginalized best fit  value of  $ z_{c}$ is considerably lower than the values reported in  Table (\ref{II}). Besides, the  behavior of decelerating parameter with redshift is shown in Fig. (\ref{Fig4}), in particular, its present-day value varies as  $-0.67<q_{0}< -0.55$ for the three cases mentioned in Table II, and all these values are in perfect
agreement with the one reported by WMAP-7 project \cite{WMAP7}.

The effective EOS of the mix is given by 
\be
\label{re5}
\omega_{\rm eff}=\frac{\beta\Omega_{\rm x}+\al\Omega_{\rm m}+\al \Omega_{\rm r}}{\sum_{i}\Omega_{\rm i}}-1. 
\ee
The  effective equation of state (EOS) for dark energy  is obtained from  Eq. (\ref{xq}), it reads 
\be
\label{re4}
\omega_{\rm effx}=\Big(\ga_{\rm x}-\frac{Q_{\rm x}}{\ro_{\rm x}} \Big)-1.
\ee
In Fig. (\ref{Fig4}) we plot the effective equation of state as a function of
redshift for the best-fit value shown in Table I,  in general,   we find that $ \omega_{\rm eff}\geq -1$ provided that  $(1-3\ga_{s})\Omega_{\rm x}+ 4(\Omega_{\rm r}+\Omega_{\rm m})  \leq 3$, as a matter of fact  its
present-day values cover the range $[-0.74, -0.64]$.  On the other hand, the  effective EOS associated to the dark energy evaluated at $z=0$, $\omega_{\rm effx}(z=0)$  varies over the range $[-1.32, -1.23]$.

\begin{center}
\begin{table}
\begin{minipage}{1\linewidth}
\scalebox{0.59}{
\begin{tabular}{|l|l|l|l|l|l|l|l|l|l|l|}
\hline
\multicolumn{11}{|c|}{Bounds for cosmological parameters}\\
\hline
$\theta_{c}$&$z_{\rm c}$ & $q(z=0)$ & $\omega_{\rm ove}(z=0)$ &$\omega_{\rm effx}(z=0)$& $\Omega_{\rm x}(z \simeq 1100)$ & $\Omega_{\rm x}(z \simeq 10^{10})$&  $\Omega_{\rm m0}$ & $\Omega_{\rm r0}$ & $\Omega_{\rm m0}/\Omega_{\rm x0}$ &$\Omega_{\rm r0}/\Omega_{\rm x0}$\\
\hline
{$I$}& $0.75$& $-0.59$& $-0.64$ & $-1.32$  &$0.26$ &$0.26$& $0.3$& $0.062$& $0.47$ & $0.09$\\
{$II$}& $0.93$& $-0.67$& $-0.74$ & $-1.23$  &$0.23$ &$0.23$& $0.3$& $0.03$& $0.44$ & $0.04$\\
{$III$}& $0.68$& $-0.55$& $-0.70$ & $-1.28$  &$0.20$ &$0.20$& $0.3$& $0.07$& $0.48$ & $0.1$\\
\hline
\end{tabular}}
\caption{\scriptsize{Derived  bounds for cosmic parameters using the best fits value of  2D C.L. obtained in  Table. (\ref{I}) by varying two cosmological parameters in three different cases.}}
\label{II}
\end{minipage}
\end{table}
\end{center}

In regard to  the behavior of  density parameters  $\Omega_{\rm x} $, $\Omega_{\rm m}$,
and $\Omega_{\rm r }$, we see that very close to $z=0$ the dark energy is the main agent that speeds up the universe, far away from $z=1$
the universe is dominated by the dark matter and at very early times the radiation component  governs the entire dynamic
of the universe around $z \simeq 10^3$[cf. Fig. (\ref{Fig4}) ]. In this point, we would like to present an appealing discussion concerning the triple cosmic coincidence problem (TCC) \cite{TCC} related to the amount of dark energy, dark matter, and radiation at present moment.   We have proposed a physical mechanism based on a phenomenological interaction among the three cosmic components, in fact, since we are working within the framework of three interacting cosmic components the aforesaid scenario seems to be a fertile arena for studying   the TCC.  As  pointed out  by Arkani-Hamed \emph{et al} in their seminal work:`` there is an era in the history of the universe where all three forms of energy, in matter, radiation and dark energy, become comparable within a few orders of magnitude''\cite{TCC}.  Here,  we have found that interaction  made it possible to have $\Omega_{\rm m0}/\Omega_{\rm x0}=0.3/0.62=0.48$ and $\Omega_{\rm r0}/\Omega_{\rm x0}=0.07/0.62=0.11$, so $\Omega_{\rm m0}/\Omega_{\rm x0} \simeq \Omega_{\rm r0}/\Omega_{\rm x0} \simeq {\cal O}(1)$, showing that the interaction implemented can be used for alleviating the TCC [see Table (\ref{II})].

Now, we explore another kind of constraint which comes form the physics at early times because this can be considered as a complementary tool for testing
our model. As is well known the fraction of dark 
energy in the recombination epoch should fulfill the bound $\Omega_{\rm ede}(z\simeq 1100)<0.1$.  Taking into account the best-fit values reported in Table (\ref{I}), we find that at  early times the dark energy  does not change much  with the redshift $z$ over the interval $[10^{3}, 10^{10}]$, in fact, the ${\rm Log}~ \Omega_{x}$ in terms of ${\rm Log~ z}$ goes from $0.64$ to $0.20$ [see Fig. (\ref{Fig4}))]. Table (\ref{II})  shows that  around $z \simeq 1100$ (recombination)  
$\Omega_{\rm x}$ can vary from $0.20$ to $0.26$. 
This excess of dark energy  requires further research  because  some signal  could arise from this early dark energy(EDE) models uncovering the nature of DE as well as their properties to high redshift, giving an invaluable guide to the physics behind the recent speed up of the universe \cite{EDE1}, \cite{hmi2}.
Regarding the values reached by $\Omega_{\rm x}$ around  the big bang nucleosynthesis (BBN) $z=10^{10}$, we find that there is a
variation from 0.20 to 0.26, so the conventional BBN processes that occurred at temperature of $1 {\rm  Mev}$ is not spoiled because the severe bound reported  for early dark energy $\Omega_{\rm x}(z\simeq 10^{10})<0.21$  is marginally fulfilled at BBN  \cite{Cyburt}. 


As is well known, dark energy dominates the whole dynamics of the universe at present  and 
there is an obvious decoupling with radiation practically. However, from a theoretical point of view,  it is reasonable to expect  that dark components can interact with other fluids of the universe substantially  in the very beginning of its evolution due to processes occurring in the early universe. For instance, dark energy interacting with neutrinos was investigated in \cite{GK}.
The framework of many interacting components could provide a more  natural  arena for studying  the stringent bounds of dark energy at recombination epoch.  There could be a signal in favor of  having dark matter exchanging energy with dark energy while  radiation is treated as a decoupled component \cite{hmi1}, \cite{hmi2} or the case where  dark matter, dark energy, and radiation exchange energy. More precisely, when the universe is filled with interacting dark sector plus a decoupled radiation term, it was found that $\Omega_{\rm x}(z\simeq 1100)=0.01$ \cite{hmi1} or $\Omega_{\rm x}(z\simeq 1100)=10^{-8}$ \cite{hmi2} but if radiation is coupled to the dark sector, the amount of dark energy is drastically reduced, giving $\Omega_{\rm x}(z\simeq 1100)\simeq {\cal{O}}(10^{-11})$ \cite{jefe2}. In our model, we have found that  the amount of early dark energy  varies in the range $[0.26; 0.20]$, so the behavior of dark energy  at recombination is  considerably  much   smoother than in the aforesaid cases \cite{hmi1}, \cite{hmi2}, \cite{jefe2}. We expect to include a (decoupled) neutrino term in Friedmann equation  to examine in more detail the dark radiation as a signature of dark energy.

\section{conclusion}

We have discussed a class of  interacting  dark matter, dark radiation, and holographic Ricci-like  dark energy model for a spatially flat FRW background.   We have coupled those components and obtained their energy densities in terms of the scale factor.

We have  examined the previous model  by constraining the cosmological parameters with the Hubble data and the well-known bounds  for dark energy at recombination era.  In the case of two-dimensional (2D) C.L., we have made three statistical constraints with the Hubble function [see Fig. (\ref{Fig1}) and Table (\ref{I})]. We have found that $\ga_{s} \leq 10^{-3} $ , so  these values  fulfill
the constraint $\ga_{s}<2/3$ for getting an accelerated phase of the universe at late times.  We find the best fit at 
$(H_{0}, \Omega_{\rm x 0})= (72.05^{+2.90}_{-2.93} {\rm km~s^{-1}\,Mpc^{-1}}, 0.637^{+0.033}_{-0.038})$ with $\chi^{2}_{\rm d.o.f}=0.81$
by using the prior $ \ga_{s}=10^{-3}$. It turned out that the  estimated values of $H_{0}$ and $\Omega_{\rm x 0}$ are in agreement with 
the standard ones reported by the WMAP-7 project \cite{WMAP7}. Besides,  we have found that $H_{0} \in [70.4, 73.81]{\rm km~s^{-1}\,Mpc^{-1}}$, 
so the estimated values are met  within $1 \sigma$ C.L.  reported
by Riess \emph{et al} \cite{H0}, to wit,  $H_{0}=(72.2 \pm 3.6){\rm km~s^{-1}\,Mpc^{-1}}$.  After having marginalized  the joint probability $P(H_{0}, \Omega_{\rm x0}, \ga_{s})$  over  $H_{0}$ [see Fig.  (\ref{Fig1})],  we saw that the marginalized   best-fit values are $(\Omega_{\rm x0}, \ga_{s})=(0.619^{+0.052}_{-0.023}, 0.001^{+0.239}_{-0.239})$  with a $\chi^{2}_{\rm d.o.f}=0.87<1$ [see Fig.  (\ref{Fig2})]. Using the best fits mentioned in Table (\ref{I}), we have numerically obtained the distance modulus of the supernova  as predicted by the theoretical model and compare with the Union 2 dataset, finding that at low redshift ($z<1.4$) there is excellent  agreement between the theoretical model and the observational data [see Fig. (\ref{Fig5})].

Regarding the derived cosmological parameters,  for instance, the transition redshift $z_{\rm c}$ turned out to be  of the order unity varying over the interval  $[0.68, 0.93]$, such values  are in agreement with $z_{\rm c}=0.69^{+0.20}_{-0.13}$ reported in \cite{Zt1}--\cite{Ztn} , and meets  within the $2 \sigma$ C.L  obtained with the supernovae (Union 2) data in \cite{Zt2}.  We also have performed a global statistical analysis using $z_{c}$ as independent parameter instead of $\Omega_{x0}$ which  lead to  $(H_{0}, z_{\rm c}, \ga_{s})=(72.05, 0.63, 10^{-3})$ along with $\chi^{2}_{\rm d.o.f}=0.86$ [see Fig. (\ref{Fig3})],  whereas the marginalized best-fit values are   $(z_{\rm c}, \ga_{s})=(0.623^{+0.039}_{-0.052}, 0.001^{+0.313}_{-0.733})$  together with  a $\chi^{2}_{\rm d.o.f}=0.86$ [see Fig. (\ref{Fig3})]. Besides, with the decelerating parameters $q(z=0) \in [-0.67, -0.55]$ for the three cases mentioned in Table (\ref{II}), all these values are  perfectly in agreement with the one reported by WMAP-7 project \cite{WMAP7} [see Fig. (\ref{Fig4})]. 

Concerning  the effective  equation of state,  we have found  that $\omega_{\rm eff}> -1$  and  its present-day values vary over the ranges  $[-0.74, -0.64]$ [see Table (\ref{II}) and  Fig. (\ref{Fig4})]. The equation of state associated with dark energy satisfies  the inequality $\omega_{\rm effx}\leq -1$. 

Besides,  we have found that  the fraction of dark radiation  at present moment $\Omega_{\rm r0}$ varies in the interval $ [0.03,  0.07]$ for the three cases mentioned in Table (\ref{II}).  The dark energy amount $\Omega_{\rm x}(z)$ governs the dynamic of the universe near $z=0$,  whereas far away from $z=1$ the universe is dominated by the fraction of  dark matter  $\Omega_{\rm m}(z)$  and at very early times the fraction of radiation $\Omega_{\rm r}(z)$ controls the entire dynamic of the universe around $z \simeq 10^3$[cf. Fig. (\ref{Fig4})].  We also have examined the triple cosmic coincidence problem \cite{TCC}  within the framework of three interacting cosmic components,  finding that  interaction used in this work provides a phenomenological  mechanism for alleviating TCC, leading to  $\Omega_{\rm m0}/\Omega_{\rm x0} \simeq \Omega_{\rm r0}/\Omega_{\rm x0} \simeq {\cal O}(1)$  [see Table (\ref{II})].

Finally,  we  have found that at  early times the dark energy  does not change much  with the redshift $z$ over the interval $[10^{3}, 10^{10}]$, in fact, the ${\rm Log}~ \Omega_{x}$ in terms of ${\rm Log~ z}$ goes from $0.64$ to $0.20$ [see Fig. (\ref{Fig4}))]. Table (\ref{II})  shows that  around $z \simeq 1100$ (recombination)  
$\Omega_{\rm x}$ can vary from $0.20$ to $0.26$.  The latter results indicate an excess of dark energy  so it  requires further research \cite{EDE1}, \cite{hmi2} , in fact   it could be related with the degeneracy presents in the equation of states of dark matter and dark radiation. In order to explore this issue in more detail, we expect to include an additional (decoupled) neutrino term in Friedmann equation; thereby,  we will seek to distinguish the radiation term coupled to dark matter, where both component share the same bare equation of state, from the decoupled neutrino term. However, it must be stressed that the values reached by $\Omega_{\rm x}$ around  the big-bang nucleosynthesis (BBN) $z=10^{10}$ vary from 0.20 to 0.26, so the conventional BBN process is not spoiled because our estimations, in most of the cases mentioned above,  fulfill the severe bound reported  for early dark energy: $\Omega_{\rm x}(z\simeq 10^{10})<0.21$ \cite{Cyburt}.  


\acknowledgments
L.P.C thanks  the University of Buenos Aires under Project No. 20020100100147 and the Consejo Nacional de Investigaciones Cient\'{\i}ficas y T\' ecnicas (CONICET) under Project PIP 114-200801-00328 for the partial support of this work during their different stages. M.G.R is partially supported by Postdoctoral Fellowship programme of  CONICET. 





\end{document}